\documentclass{jaa}
\usepackage{threeparttable}
\usepackage{amsmath}
\usepackage[utf8]{inputenc}
\usepackage{longtable}
\usepackage{graphicx}
\usepackage{natbib}
\usepackage{multirow}
\usepackage[Table,xcdraw]{xcolor}
\usepackage{rotating}
\newcommand{\apj}{Astrophysical Journal}

\newcommand{\aap}{Astronomy and Astrophysics}

\usepackage{hyperref}
\hypersetup{backref=true,
    pagebackref=true,
    hyperindex=true,
    colorlinks=true,
    breaklinks=true,
    urlcolor= black,
    linkcolor= blue,
    bookmarks=true,
    bookmarksopen=false,
    filecolor=black,
    citecolor=blue,
    linkbordercolor=blue
}

\begin{document}\sloppy

\title{Core-collapse supernova from a possible progenitor star of 100\,M$_{\odot}$}


\author{Amar Aryan\textsuperscript{1,2}, Shashi Bhushan Pandey\textsuperscript{1,2}, Abhay Pratap Yadav\textsuperscript{3}, Amit Kumar\textsuperscript{1,4}, Rahul Gupta\textsuperscript{1,2}, and Sugriva Nath Tiwari\textsuperscript{2}}

\affilOne{\textsuperscript{1}Aryabhatta research institute of observational sciences (ARIES), Nainital, Uttarakhand, India, \\}

\affilTwo{\textsuperscript{2}Department of Physics, Deen Dayal Upadhyaya Gorakhpur University, Gorakhpur-273009, India \\}

\affilThree{\textsuperscript{3}Department of Physics and Astronomy, National Institute of Technology Rourkela-769008, Odisha, India\\}

\affilFour{\textsuperscript{4}School of Studies in Physics and Astrophysics, Pt. Ravishankar Shukla University, Chattisgarh 492010, India\\}


\twocolumn[{

\maketitle

\corres{amararyan941@gmail.com, amar@aries.res.in}

\msinfo{---}{---}

\begin{abstract}
{In this work, we study the synthetic explosions of a massive star. We take a 100\,M$_{\odot}$ zero--age main--sequence (ZAMS) star and evolve it until the onset of core-collapse using {\tt MESA}. Then, the resulting star model is exploded using the publicly available stellar explosion code, {\tt STELLA}. The outputs of {\tt STELLA} calculations provide us the bolometric light curve and photospheric velocity evolution along with other physical properties of the underlying supernova. In this paper, the effects of having large Hydrogen-envelope on the supernova light curve have been explored. We also explore the effects of the presence of different amounts of nickel mass and the effect of changing the explosion energy of the resulting supernovae from such heavy progenitors, on their bolometric light curves and photospheric velocities. }

\end{abstract}

\keywords{Supernovae---Black holes---Neutron stars---{\tt MESA}---{\tt STELLA}}

}]


\doinum{}
\artcitid{\#\#\#\#}
\volnum{000}
\year{0000}
\pgrange{1--}
\setcounter{page}{1}
\lp{9}

\section{Introduction}
\label{Intro}

Supernovae (SNe) are bright and extremely powerful explosions that mark the death of stars, either massive ones or a white dwarf. Based on their death throes, SNe are broadly classified into two categories, namely, thermonuclear SNe and core-collapse SNe (CCSNe). The thermonuclear SNe are thought to be the result of the explosion of a carbon-oxygen white dwarf in a binary system as it exceeds the Chandrasekhar mass limit, either due to accretion from a companion \citep[][]{Whelan1973} or
mergers \citep[][]{Maoz2014,VanRossum2016,Konyves2020}. The thermonuclear SNe are the end results of the stellar evolution of the low mass stars ($\lesssim 8\,M_{\odot}$). The CCSNe are the final destinations of massive stars ($\gtrsim$ 8--10\,$M_\odot$;  e.g., \citealt[][]{Garry2004, Woosley2005, Groh2017}), resulting from their core-collapse due to the exhaustion of the nuclear fuel in their core. Underlying mechanisms of core-collapse of massive stars are still not understood well. One popular mechanism is the neutrino--driven outflow \citep[][]{Muller2017} but others mechanisms have also been proposed (e.g., magnetorotational mechanism of the explosion of CCSNe as discussed in \citet[][]{Bisnovatyi2018}). In many cases, the CCSNe explosions are not spherically symmetrical. As a particular case, studies by \citet[][]{Couch2009} indicate that the aspherical CCSNe in red supergiants are powered by non- relativistic Jets. Further, \citet[][]{Piran2019} mentions that gamma ray bursts (GRBs) that accompany rare and powerful CCSNe (popularly known as 'hypernovae') involve the association of relativistic jets. However \citet[][]{Piran2019} also mentions about the co-existence of SNe and relativistic jets in the long GRBs, but the relationship between the mechanisms that drives the hypernova and the jet is still unknown. 

The thermonuclear SNe and CCSNe are observationally divided into two main classes based on the presence or absence of Hydrogen (H)--features in their spectra. SNe showing prominent H--features in their spectra are classified as Type II, while Type I SNe cease to show strong H--features in their spectra. These two classes are further divided into various subclasses. 
Among Type I SNe, the Type Ia SNe display a strong silicon absorption line near the maximum luminosity; the Type Ib SNe exhibit prominent Helium (He)--features in their spectra, whereas Type Ic SNe show neither H nor He obvious features. Prominent features of intermediate-mass elements such as O, Mg, and Ca are also seen in Type Ib and Type Ic SNe spectra.

The Type II SNe are historically classified into Type IIP (plateau) and Type IIL (linear) based on the shapes of their light curve \citep[][]{Barbon1979, Filippenko1997}. Extensive reviews of various types of CCSNe and criteria used to categorise them are provided by (among others) \citet[][]{Filippenko1997} and \citet[][]{Gal-Yam16}. A very interesting class of SNe known as Type IIb SNe, also exists. These Type IIb SNe form a transition class of objects that are supposed to link SNe~II and SNe~Ib \citep[][]{Filippenko1988, Filippenko1993, Smartt2009}. The early-phase spectra of SNe~IIb display prominent H--features, while unambiguous He--features appear after a few weeks \citep[][]{Filippenko1997}.

During the last stages of their lives, the stars are fully evolved and contain mostly intermediate (eg. Si, Mg etc.) to high mass elements (eg. Fe, Ni, Co etc.). Upon their deaths through massive explosions, these elements are thrown into the vast space, thus these catastrophic events are responsible for the enrichment of our universe. SNe are also responsible for the birth of compact objects like neutron stars and black holes.

In this work we take a 100\,M$_{\odot}$ ZAMS star with metallicity, Z = 0.01 and evolve it up to the stage of core-collapse and finally, the synthetic explosions are simulated.
Depending on the mass loss rate, rotation and even on metallicity, such a heavy progenitor could result into various transients including pair instability supernova (PISN), pulsational pair instability supernova (PPISN), Type IIp, Type IIn, Type Ibn and Type Icn SNe also \citep[][]{Sukhbold2016,Woosley2017}. The numerical simulations of such heavy progenitors are always a challenging task. With the mass loss and rotation included, the simulations are further complicated. The type IIp SNe progenitors suffer minimal  mass loss ( as they retain almost all of their H-envelope ), and possess (very) low rotation. Thus, in order to keep things simple yet close to reality, we try to model a massive ZAMS progenitor that can result into a type IIp like SN. One very recent study by \citet[][]{Nicholl2020} indicates a total (Supernova ejecta + CSM) mass likely exceeding 50--100\,M$_{\odot}$ for a Type II superluminous SN~2016aps. Attempts are underway to simulate such massive progenitors with much lower metallicities close to zero and even models with zero metallicity to understand the internal structure and resulting transients in future. Although a few studies have been performed in this direction by (among others) \citet[][]{Ohkubo2006}, such studies are going to be extremely useful to understand the final fates of massive primordial stars. The primary reason to choose a 100\,M$_{\odot}$ ZAMS progenitor is that the mass ranges including such heavy progenitors are least explored. Only a handful of studies \citep[eg.][]{Sukhbold2016,Woosley2017} have been performed which discuss the evolution and final fates of such massive zero-age main-sequence progenitors. This study is a preliminary step to enhance our present understanding of CCSNe having very high mass progenitors. It also demonstrates the suitability of {\tt MESA} and {\tt STELLA} to study such CCSNe. Further, as mentioned above, a recent study by \citet[][]{Nicholl2020} indicates a total (SN ejecta + CSM) mass exceeding 50--100 M$_{\odot}$, so our study is also going to be important to look into the progenitors of such events. Moreover, in the coming TMT\footnote{\url{https://www.tmt.org/}}, ELT\footnote{\url{https://elt.eso.org/}} and other large telescopes, such massive stars are likely to be discovered, thus it will always be an interesting task to compare the simulations with the actual observations. All the analyses performed in this work make use of publicly available tools. A very nice review has been provided (among others) in \citet[][]{Aryan2021b} depicting the usefulness of publicly available tools and how these tools can be boon to the transient community.    

This paper has been divided into six sections. The basic assumptions and methods of modeling the 100\,M$_{\odot}$ progenitor have been discussed in section~\ref{Models}, thereafter the properties and parameters to synthetically simulate the explosions are discussed in section~\ref{Explosion}. The results of synthetic explosions have been discussed in section~\ref{Results}.  Later, we discuss the major outcomes of our studies in section~\ref{Discussion}, and provide our concluding remarks in section~\ref{Conclusion}.

\section{Progenitor modeling using {\tt MESA}} 
\label{Models}
{\tt MESA} \citep[][] {Paxton2011,Paxton2013,Paxton2015,Paxton2018,Paxton2019} is an openly available one dimensional stellar evolution code with a suite of open source, rich, efficient, thread-safe libraries for a wide range of applications in computational stellar astrophysics. It is a very useful tool to study pulsations in stars, accretion onto a Neutron star, the stages of stellar evolutions causing various types of SNe and many other important astrophysical phenomena. Here we briefly mention the {\tt MESA} settings and assumptions for our calculations.

\begin{figure*}[!t]
  \includegraphics[height=12cm,width=17.5cm]{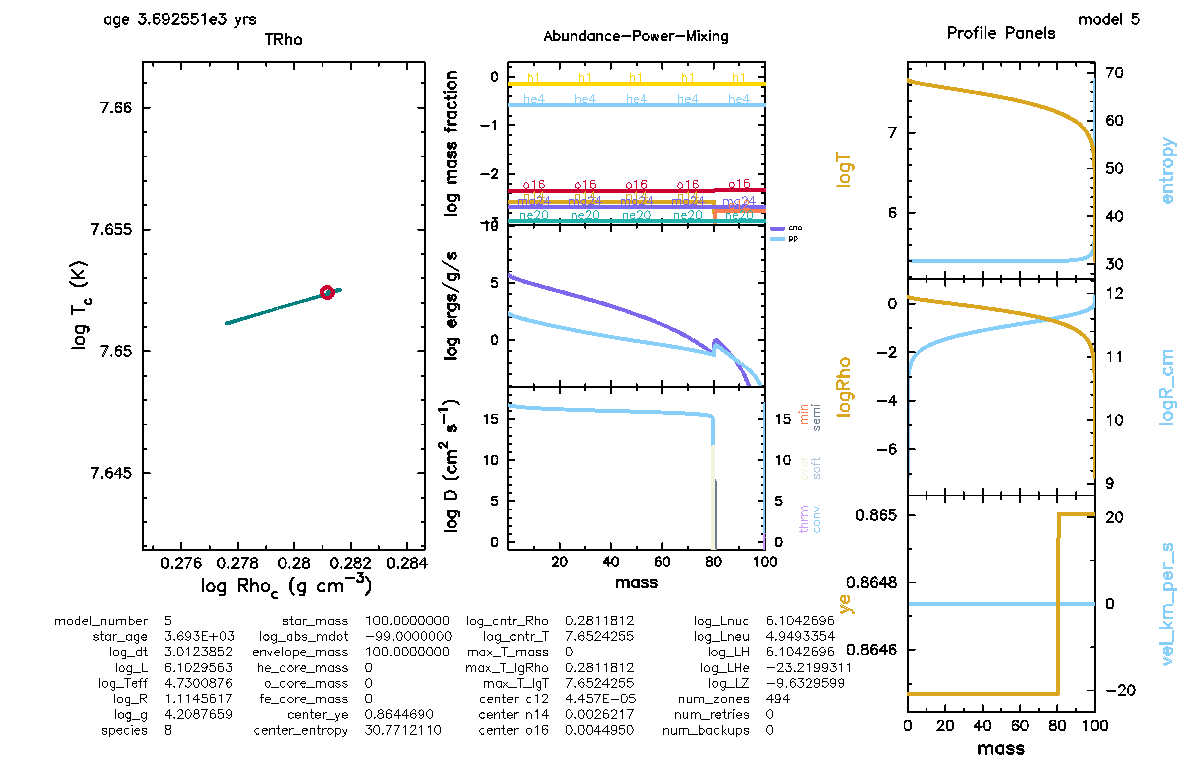}
  \caption{The physical condition and chemical engineering of our model when it has just arrived on the ZAMS. This figure is one of the many figures generated by {\tt MESA} while the evolution of the models through various evolutionary phases.}
  \label{fig:mesa1}
\end{figure*}

In this work, we have taken a 100 M$_{\odot}$ ZAMS progenitor and evolved it upto the onset of core-collapse using {\tt MESA} (version r11701). For this purpose, we used the default {\tt example\_make\_pre\_ccsn} directory available in {\tt MESA}. Most of the settings closely follow \citet[][]{Aryan2021} and  \citet[][]{Pandey2021} with only a few minor deviations. We do not consider rotation and assume an initial metallicity of $Z = 0.01$. Convection is modelled using the mixing theory of \citet[][]{Henyey1965} by adopting the Ledoux criterion. We set the mixing-length parameters to $\alpha = 3.5$ in the region where the mass fraction of hydrogen is greater than 0.5, and set it to 1.5 in the other regions. Semi-convection is modelled following \citet[][]{Langer1985} with an efficiency parameter of $\alpha_{\mathrm{sc}} = 0.01$. For the thermohaline mixing, we follow \citet[][]{Kippenhahn1980}, and set the efficiency parameter as $\alpha_{\mathrm{th}} = 2.0$. We model the convective overshooting with the diffusive approach of \citet[][]{Herwig2000}, with $f= 0.004$ and $f_0 = 0.001$ for all the convective core and shells. We do not consider the presence of any strong stellar winds and hence consider no progenitor moss loss prior to the core-collapse. 

\begin{figure*}[!t]
  \includegraphics[height=12cm,width=17.5cm]{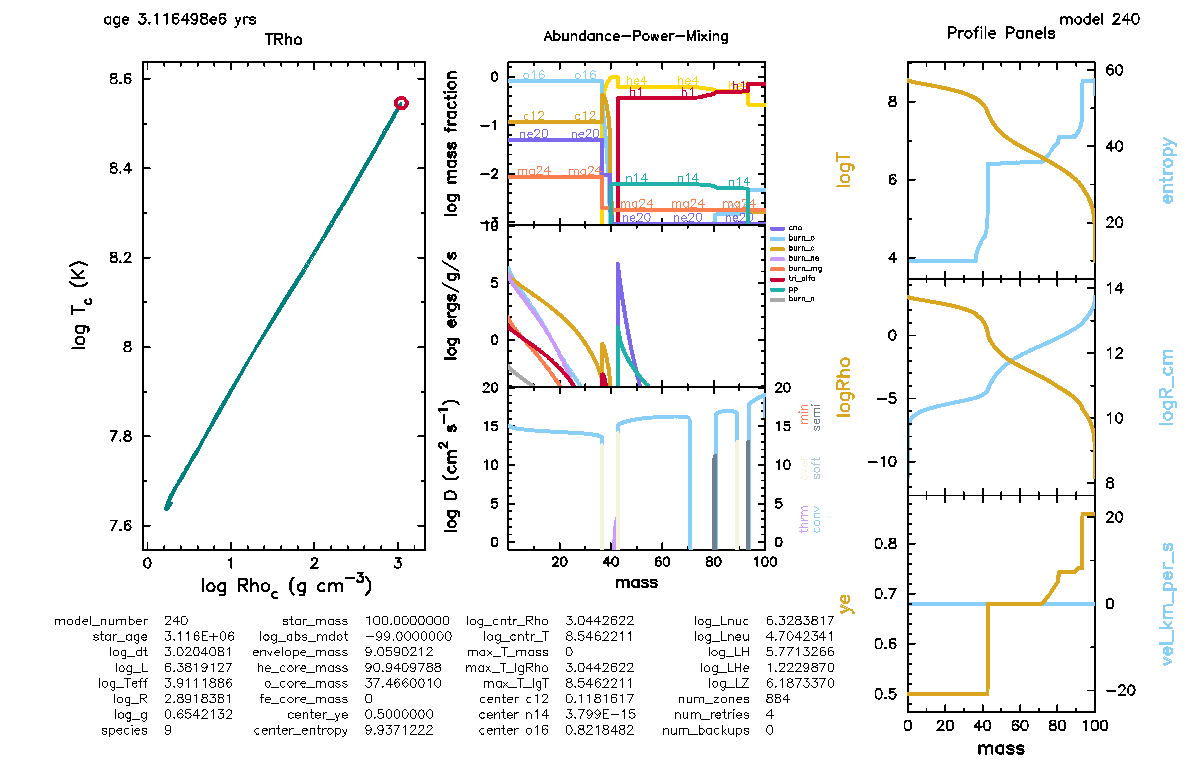}
  \caption{Same as fig.~\ref{fig:mesa1}, but at the stage when the model has evolved through the main sequence and has become red supergiant.}
  \label{fig:mesa2}
\end{figure*}

\begin{figure*}[!t]
  \includegraphics[height=12cm,width=17.0cm]{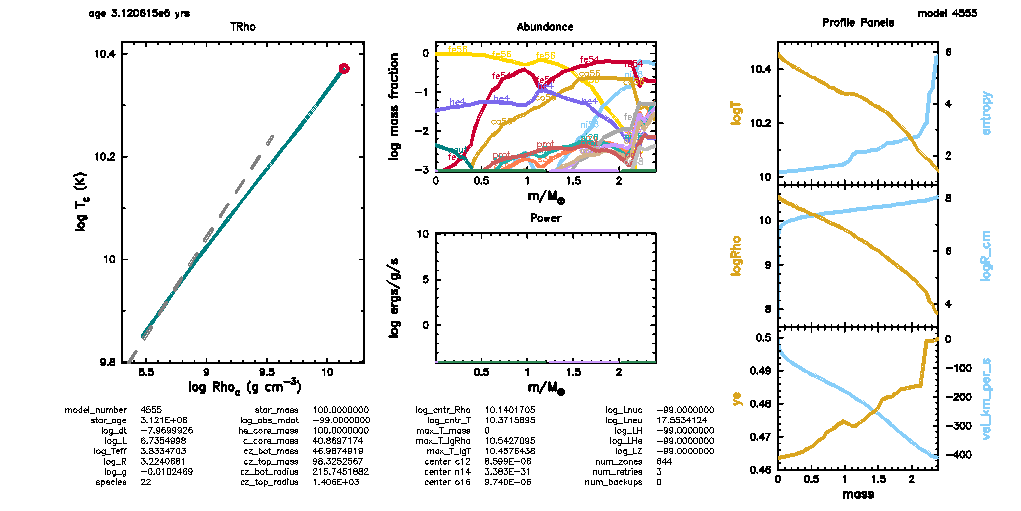}
  \caption{A glimpse of the physical and chemical structure of our model star just when the Iron-core infall has been achieved. We could see that the core temperature and density have reached in excess of $10^{10}\,K$ and $10^{10}\,g\,cm^{-3}$ respectively, that are supposed to be the perfect conditions for the core-collapse.}
  \label{fig:mesa3}
\end{figure*}

Further, we briefly explain the evolution of our models through the various stages. Landing on the ZAMS, our model evolves through the main-sequence,  becomes supergiant and finally approaches its death by developing an inert Fe-core in its center. Figure~\ref{fig:mesa1} displays the physical condition and chemical engineering of our model when it has just landed on the main sequence. The arrival of our model on the main sequence is marked by the control parameter {\tt Lnuc\_div\_L\_zams\_limit}, which checks the ratio of luminosity due to nuclear reactions and total luminosity at ZAMS. Once this ratio is $\sim$ 0.8, the onset of the main-sequence phase is marked. The leftmost panel of figure~\ref{fig:mesa1} shows the variation of the core temperature (log\,T$_{c}$)  and core density (log\,Rho$_{c}$) as the model evolves in time. On the beginning of main-sequence, the log\,T$_{c}$ is $\sim$ 7.65\,K and log\,Rho$_{c}$ is $\sim 0.281\,g\,cm^{-3}$. The top plot in the middle panel of figure~\ref{fig:mesa1} displays the mass fraction of various elements present in the star. We can see that as the model has just arrived on the main-sequence, the fractions of H and He are much larger than any other heavy metals depending on the metallicity of our model. The middle plot in the middle panel of figure~\ref{fig:mesa1} shows the specific luminosity corresponding to the various processes going inside the star. It can be seen that only two processes have kicked in as our model star is just beginning its journey through the main-sequence. The bottom plot in the middle panel shows the variation of the kinematic viscosity inside our model star. Now the rightmost panel of figure~\ref{fig:mesa1} shows various physical properties of our model star. Here, the top most plot shows the variation of temperature and entropy with the mass coordinate, the middle plot shows the variation of density (logRho) and radius (logR\_cm) of the model star with mass coordinate and finally the bottom plot shows the variation of baryon fraction (ye) and convectional velocity (vel\_km\_per\_s) inside the star.        

The figure~\ref{fig:mesa2} shows similar properties as in figure~\ref{fig:mesa1}, but at a stage when our model star has evolved through main-sequence and has become a supergiant. The leftmost panel in figure~\ref{fig:mesa2} shows that now the core temperature and density have risen considerably. Here, the top plot in the middle panel of figure~\ref{fig:mesa2} indicates the presence of heavier elements near the core while H and He are present in envelope towards the higher mass coordinates. The middle plot in the middle panel of figure~\ref{fig:mesa2} shows that multiple energy generation processes have kicked in, while the bottom plot shows the further variation of kinematic viscosity inside our model star. Now, the rightmost panel of figure~\ref{fig:mesa2} depicts the variation of various physical parameters. From the radius plot, we can see that the model star has swollen upto $\sim 10^{14}\,cm$ which is roughly around $1400\,R_{\odot}$ while from the temperature plot, we can see that the temperature (logT plot) near the surface is $\sim 10000\,$K. Such radius and temperature clearly indicate that our model is living the supergiant phase. 

Finally, figure~\ref{fig:mesa3} shows similar properties but at a stage when the core of our model star is about to collapse. The leftmost panel of figure~\ref{fig:mesa3} shows the log\,T$_{c}$ and log\,Rho$_{c}$ have reached in excess of $10^{10}\,$K and $10^{10}\,gm\,cm^{-3}$ respectively. Such core temperatures and core densities are thought to be the perfect conditions for the core to undergo collapse. Thus, the core of our model star is about to collapse. The mass fraction and specific luminosity plots in the middle panel of figure~\ref{fig:mesa3} also depict perfect conditions for the core to collapse. In the mass fraction plot, a very high fraction of $^{56}$Fe is seen which indicates that the model star has developed an inert Fe-core, also in the specific luminosity plot, the absence of any energy generation process is seen. The plots in the rightmost panels of figure~\ref{fig:mesa3} show the physical properties of the inner--core regions. All these plots indicate the arrival of the stage of core-collapse.

Post core-collapse, our model is evolved from a few seconds after the central explosion triggered by core-collapse to a time just before the outgoing shock reaches to the stellar surface as discussed in \citet[][]{Paxton2018}. For this purpose the default directory, {\tt example\_ccsn\_IIp} from {\tt MESA}  is used.

\begin{figure*}
\centering
    \includegraphics[height=8.0cm,width=8.5cm,angle=0]{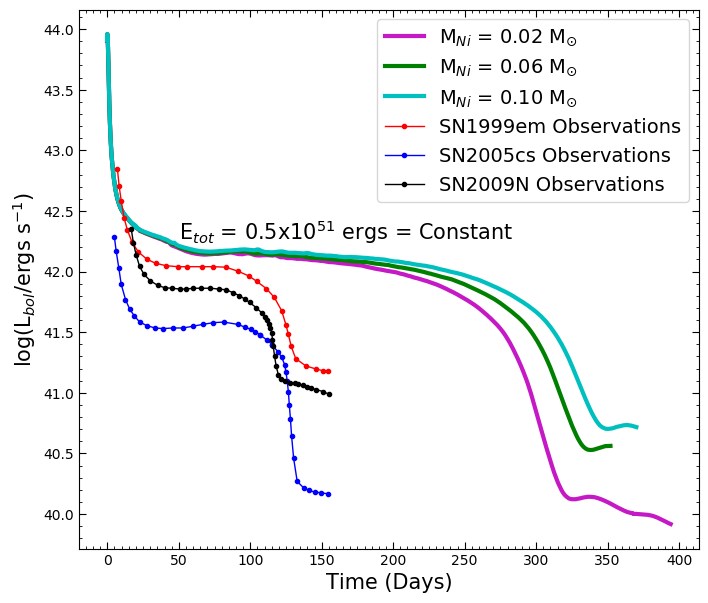}
    \includegraphics[height=8.0cm,width=8.5cm,angle=0]{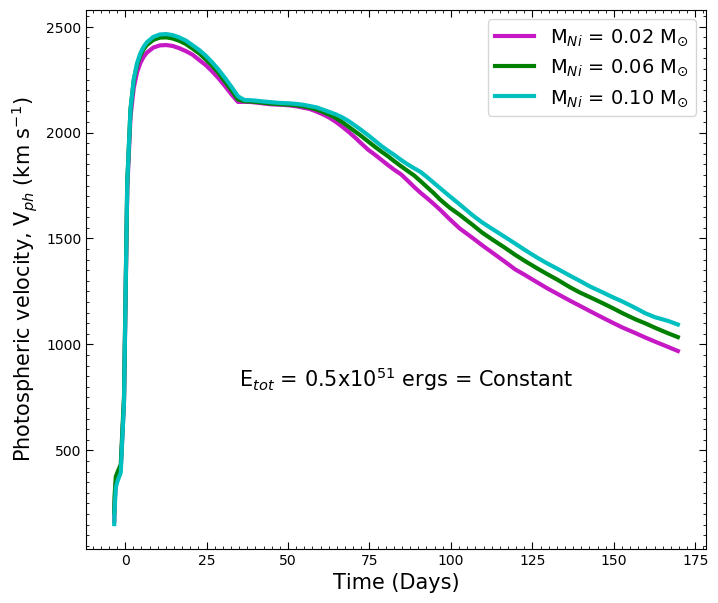}
   \caption{ Results of the synthetic explosions simulated by {\tt STELLA}, for comparison of plateau phase, the bolometric light curves of three normal type IIP SNe, namely SN1999em, SN2005cs, and SN2009N have also been plotted. The explosion energy has been kept fixed at $0.5\times10^{51}\,$ergs and the nickel mass is varied to 0.10\,M$_{\odot}$, 0.06\,M$_{\odot}$, and 0.02\,M$_{\odot}$. {\em Left:} The behaviour of bolometric light curve by the variation of nickel mass with the explosion energy kept constant has been displayed. The duration of the plateau increases with the increase in nickel mass. {\em Right:} The effect on the photospheric velocity when the nickel mass is varied and the explosion energy is kept constant. Minimal changes are seen on the photospheric velocities. }
    \label{fig:stella1}
\end{figure*}
\begin{figure*}
\centering
    \includegraphics[height=8.0cm,width=8.5cm,angle=0]{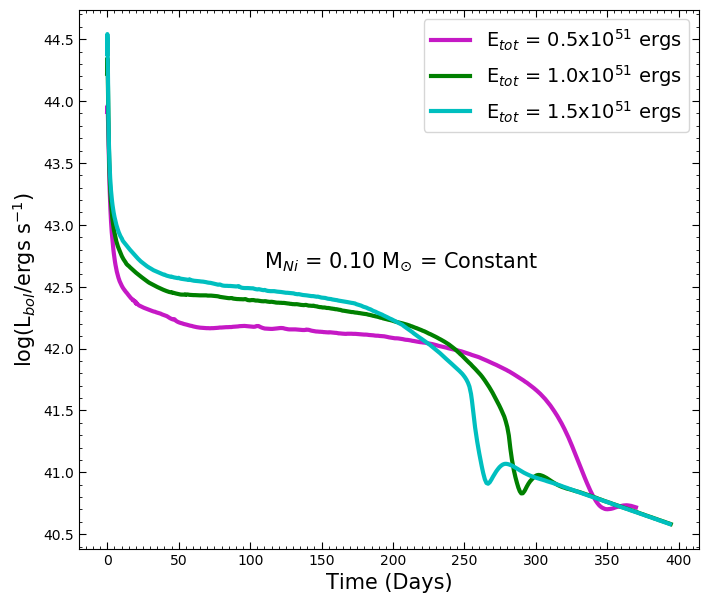}
    \includegraphics[height=8.0cm,width=8.5cm,angle=0]{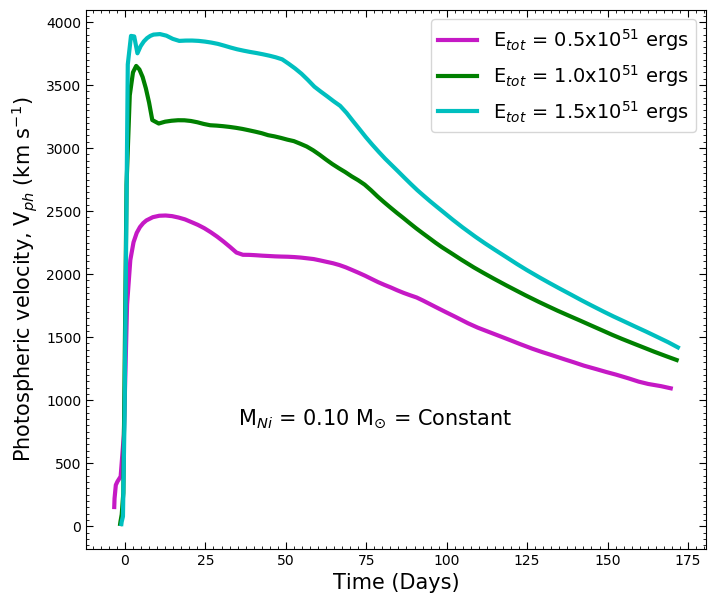}
    \caption{ Results of the synthetic explosions simulated by {\tt STELLA}. The nickel mass has been kept fixed to 0.10\,M$_{\odot}$ and the explosion energy is varied to $0.5\times10^{51}\,$ergs, $1.0\times10^{51}\,$ergs and, $1.5\times10^{51}\,$ergs. {\em Left:} The behaviour of bolometric light curve by the variation of explosion energy with the nickel mass kept constant has been displayed. The overall plateau luminosity increases with increase in explosion energy, while the plateau duration decreases with increase in explosion energy. {\em Right:} The effect on the photospheric velocity when the explosion energy is varied and the nickel mass is kept constant. The photospheric velocities increase with increase in explosion energy.}
    \label{fig:stella2}
\end{figure*}

\section{Synthetic explosions using {\tt STELLA}} \label{finding}
\label{Explosion}
The output of {\tt MESA} when the shock has reached just before the stellar surface, is fed as input to {\tt STELLA} \citep[][]{Blinnikov1998, Blinnikov2000, Blinnikov2006} which is a hydrodynamical code used to simulate the energetic core-collapse supernova explosions by evolving the model through shock-breakout. It solves the radiative transfer equations in the intensity momentum approximation in each frequency bin. In {\tt MESA}, {\tt STELLA} is  run using 40 frequency groups, which is enough to produce spectral energy distribution, but  insufficient to produce spectra. The opacity is computed based over 153,000 spectral lines from \citet{Kuruz1995} and \citet{Verner1996}. To calculate the overall opacity, the code uses 16 species, which include H, He, C, N, O, Ne, Na, Mg, Al, Si, S, Ar, Ca, a sum of stable Fe and radioactive $^{56}$Co, stable Ni and radioactive $^{56}$Ni. The expansion opacity formalism from \citet{Eastman1993} is used to compute the line opacity by taking into account the effect of high velocity gradients. The photo-ionization, electron scattering and free-free absorption are also the sources of opacity that are included. Local thermodynamic equilibrium (LTE) is assumed in the plasma so that the Boltzmann-Saha distribution for ionization and level populations can be used. No nuclear networks except radioactive decay of $^{56}$Ni to $^{56}$Co which further decays to $^{56}$Fe, are used. The energy from such$^{56}$Ni and $^{56}$Co radioactive decay is deposited into positrons and gamma-photons. This energy is treated in a one-group transport approximation as suggested by \citet{Swartz1995}.

In {\tt STELLA}, a Lagrangian co--moving grid is used to solve the 1-dimensional equations for mass, momentum and total energy. The artificial viscosity comprising of the standard von Neumann artificial viscous pressure is used for stabilizing solution \citep{Von1950} and a so-called cold artificial viscosity is used to smear shocks \citep{Blinnikov1996}. So, {\tt STELLA} enables one to properly compute the shock propagation along the ejecta and the shock breakout event. The coupled equations of radiation hydrodynamics are solved through an implicit high-order predictor–corrector procedure based on the methods of \citet{Brayton1972} ( more details in \citep{Blinnikov1996} and \citep{Stabrowski1997}).

Our {\tt STELLA} calculations include the synthetic explosions of a 100 M$_{\odot}$ ZAMS progenitor by assuming a set of explosion energies (E$_{tot}$) and nickel masses (M$_{Ni}$). The radioactive decay of $^{56}$Ni and $^{56}$Co powering mechanism has been employed in {\tt STELLA}.  We explored the effect on the bolometric light curve and photospheric velocity by changing one variable out of (E$_{tot}$) and (M$_{Ni}$) and keeping the other constant. In the first case, the explosion energy is kept constant to $0.5\times10^{51}\,$ergs while the nickel mass is changed to 0.10\,M$_{\odot}$, 0.06\,M$_{\odot}$, and 0.02\,M$_{\odot}$. The masses of the central remnants for these cases were $\sim$41\,M$_{\odot}$. For the second case, the nickel mass is kept constant to 0.10\,M$_{\odot}$ and the explosion energy is varied to $0.5\times10^{51}\,$ergs, $1.0\times10^{51}\,$ergs and, $1.5\times10^{51}\,$ergs. The masses of the central remnants were different corresponding to different explosion energies. Corresponding to the explosion energies of $0.5\times10^{51}\,$ergs, $1.0\times10^{51}\,$ergs and, $1.5\times10^{51}\,$ergs, the masses of the central remnants were $\sim$\,41\,M$_{\odot}$, 38\,M$_{\odot}$ and 31\,M$_{\odot}$ respectively.

\section{Results}
\label{Results}
We evolved a 100 M$_{\odot}$ ZAMS progenitor upto the onset of core-collapse using {\tt MESA} and the synthetic explosions of the model are simulated using {\tt STELLA}. The effects of variations of the explosion energy and nickel mass on the bolometric luminosity and photospheric velocity is explored. The left and right panels of figure~\ref{fig:stella1} show the effect of variation of nickel mass by keeping the explosion energy fixed on the bolometric light curve and photospheric velocity evolutions respectively. First, an increased duration of plateau is noticed in comparison to normal type IIP SNe. Such an increased duration of plateau is attributed to the huge, inflated H--envelope of the model star. For comparison purposes, we have also plotted three normal type IIP SNe, namely, SN~1999em, SN~2005cs, and SN~2009N (the light curves have been taken from \citet[][]{Paxton2019}) in the left panel of figure~\ref{fig:stella1}. From modeling their bolometric light curves, \citet[][]{Paxton2019} found that their progenitors have masses in the range 13\,M$_{\odot}$ to 19\, M$_{\odot}$. These SNe have much lower extent of H-envelope as compared to our 100\,M$_{\odot}$, so they have much lower plateau phase. Thus, the H-envelope of a supernova progenitor plays a very important role in determining the length of the plateau phase. 

Further, it is noticed that with the increase in the Nickel mass, the duration of plateau is also increased (left panel of figure~\ref{fig:stella1}). Due to the increased amount of Nickel, there is an increase in the energy deposition in the expanding ejecta by the gamma--rays produced by the radioactive decay of $^{56}$Ni and $^{56}$Co, which prolongs the plateau duration. Since the explosion energy is kept fixed, the ejecta move with nearly similar velocity for each case of different Nickel mass thus, minimal variation is seen on the photospheric velocities ( right panel of figure~\ref{fig:stella1}).

Figure~\ref{fig:stella2} shows the effect of variation of explosion energy by keeping Nickel mass constant, on the bolometric luminosity and photospheric velocity evolutions. In the left panel of figure~\ref{fig:stella2}, we can see two prominent effects when the nickel mass is kept fixed and the explosion energy is varied; first, the overall increase in the luminosity of the plateau phase and second, the decrease in the plateau duration with increase in explosion energy. Due to the increase in the explosion energy, the supernova is intrinsically bright and powerful that is why the intrinsic brightness during the plateau phase increases with the increase in explosion energy. Also, higher explosion energy implies increase in the kinetic energy of ejecta, as a result, the supernova expands and cools faster. Thus the re-ionizing front that is responsible for the plateau is extinguished early and the duration of plateau decreases. With increase in the explosion energy, the kinetic energy of ejecta also increases resulting in faster moving ejecta which is clearly evident in the right panel of figure~\ref{fig:stella2}.

\section{Discussion}
\label{Discussion}

In this work, a 100 M$_{\odot}$ ZAMS progenitor model has been evolved from ZAMS upto the stage where it can undergo core-collapse with the help of publicly available tool {\tt MESA}. The output of {\tt MESA} is fed as input to {\tt STELLA}, which simulates the synthetic explosions. Based on the mass loss rate, rotation and metallicity, such a heavy progenitor could result into various transients including PISN, PPISN, Type IIp, Type IIn, Type Ibn and Type Icn also, but for our case we explored the possible outcomes when such heavy ZAMS mass progenitor undergoes a Type IIP like CCSNe by displaying huge plateau in the light curves. Such heavy stars under prevailing physical circumstances might lead towards exploding as superluminous SNe but analysis of such type of explosions is beyond scope of this paper and would be discussed separately in the coming future. In the following points, we briefly summarize our present work : 

1) A 100 M$_{\odot}$ ZAMS progenitor model has  much larger H-envelope compared to the typical progenitors of type IIP SNe, due to which an extended duration of plateau is seen.

2) The effect of variation of Nickel mass is explored. On increasing the Nickel mass while keeping the explosion energy constant, the duration of the plateau in the light curve is increased. This behaviour due to increase in Nickel mass is attributed to the increase in the energy deposition in the expanding ejecta by the gamma--rays that are produced by the radioactive decay of $^{56}$Ni and $^{56}$Co. Also, very minor changes are seen in the photospheric velocity evolution. This behaviour of photospheric velocity is attributed to the constant explosion energy. Due to a constant energy of explosion, the kinetic energy of the ejecta having different Nickel masses is nearly similar, resulting in nearly similar photospheric velocities.     

3) The effect of variation of explosion energy is also explored. On increasing the explosion energy and keeping the Nickel mass constant, the overall plateau luminosity increases but the duration of the plateau decreases. Such type of behaviour is attributed to the supernova becoming intrinsically bright and powerful with the increase in explosion energy. Due to the supernova becoming intrinsically bright, the luminosity of the plateau phase is increased. Also, with the increase in the explosion energy, the supernova expands and cools faster. Because of this, the re-ionizing front that is responsible for the creation of plateau gets extinguished faster and the plateau duration decreases.   

\section{Conclusion}
\label{Conclusion}
We could simulate synthetic core-collapse supernova explosions resulting from a 100 M$_{\odot}$ ZAMS progenitor. As discussed earlier, based on the mass loss rate, rotation and metallicity, such a heavy progenitor could result into various transients including PISN, PPISN, Type IIp, Type IIn, Type Ibn and Type Icn also. In this work, we explored the outcomes of the synthetic explosions from a 100 M$_{\odot}$ ZAMS progenitor resulting into Type IIP like SNe. The presence of a very large H-envelope prior to the supernova explosion resulted in an extremely large plateau for the SNe. The effects of the variation of explosion energy and nickel mass on the bolometric luminosity and photospheric velocity were also explored.

\section*{Acknowledgements}
We are thankful to the anonymous referee for providing valuable comments. A.A. acknowledges funds and assistance provided by the Council of Scientific \& Industrial Research (CSIR), India with file no. 09/948(0003)/2020-EMR-I. A.A., S.B.P. and R.G. acknowledge BRICS grant DST/IMRCD/BRICS/Pilotcall/ProFCheap/2017(G). We are thankful to the {\tt MESA} troubleshooting team for their constant support and guidance. 




\begin{thebibliography}{}

\bibitem[\protect\citeauthoryear{Aryan et al.}{2021a}]{Aryan2021} Aryan, A., Pandey, S. B., Zheng, W., Filippenko, A., V., Vinko, J. et al.\ 2021, MNRAS, 505, 2530

\bibitem[\protect\citeauthoryear{Aryan et al.}{2021b}]{Aryan2021b} Aryan, A., Pandey, S. B., Kumar, A., Gupta, R., Castro-Tirado, A. J. et al. \ 2021, RMxAC, 53, 215–224

\bibitem[\protect\citeauthoryear{Barbon, Ciatti, \& Rosino}{1979}]{Barbon1979} Barbon R., Ciatti F., Rosino L.\, 1979,  A\&A, 72, 287

\bibitem[\protect\citeauthoryear{Bisnovatyi-Kogan, Moiseenko, \& Ardelyan}{2018}]{Bisnovatyi2018} Bisnovatyi-Kogan, G. S., Moiseenko, S. G. \&, Ardelyan, N. V. \ 2018, PAN, 81, 266

\bibitem[\protect\citeauthoryear{Blinnikov \& Panov}{1996}]{Blinnikov1996} Blinnikov S. I., \& Panov I. V. 1996, Astron. Lett., 22, 39

\bibitem[\protect\citeauthoryear{Blinnikov et al.}{1998}]{Blinnikov1998} Blinnikov, S. I., Eastman, R., Bartunov, O. S., Popolitov, V. A., \& Woosley, S. E. 1998, ApJ, 496, 454

\bibitem[\protect\citeauthoryear{Blinnikov et al.}{2000}]{Blinnikov2000} Blinnikov, S., Lundqvist, P., Bartunov, O., Nomoto, K., \& Iwamoto, K. 2000, ApJ, 532, 1132

\bibitem[\protect\citeauthoryear{Blinnikov et al.}{2006}]{Blinnikov2006} Blinnikov, S. I., Röpke, F. K., Sorokina, E. I., et al. 2006, A\&A, 453, 229

\bibitem[\protect\citeauthoryear{Brayton et al.}{1972}]{Brayton1972} Brayton R. K., Gustavson F. G., Hatchel G. D. 1972, Proc. IEEE   

\bibitem[\protect\citeauthoryear{Cough, Wheeler \& Milosavljevic}{2009}]{Couch2009} Couch, S. M., Wheeler, J. C., \& Milosavljevi\'c, M. \ 2009, ApJ, 696, 953 

\bibitem[\protect\citeauthoryear{Eastman \& Pinto} {1993}]{Eastman1993}  Eastman, R. G., \& Pinto, P. A. 1993, ApJ, 412, 731

\bibitem[\protect\citeauthoryear{Fan} {2001}]{Fan2001} Fan, X., Narayanan, V., K., Lupton, R. H., Strauss, M., A., Knapp, G. R. et al.\ 2001, AJ, 122, 2833 


\bibitem[\protect\citeauthoryear{Filippenko}{1988}]{Filippenko1988} Filippenko, A. V.\ 1988, AJ, 96, 1941

\bibitem[\protect\citeauthoryear{Filippenko et al.}{1993}]{Filippenko1993} Filippenko, A.~V., Matheson T., Ho L. C.\ 1993, ApJL, 481, L89

\bibitem[\protect\citeauthoryear{Filippenko}{1997}]{Filippenko1997} Filippenko, A. V.\ 1997, ARA\&A, 35, 309

\bibitem[{Gal-Yam(2016)}]{Gal-Yam16} Gal-Yam, A.\ 2017, in Handbook of Supernovae (Springer International
  Publishing), 1

\bibitem[\protect\citeauthoryear{Garry}{2004}]{Garry2004} Garry., G.\ 2004, Science, Vol. 304, Issue 5679, pp. 1915-1916, DOI: 10.1126/science.1100370 

\bibitem[\protect\citeauthoryear{Groh}{2017}]{Groh2017} Groh Jose H.\ 2017, Predicting the nature of supernova progenitors, Phil. Trans. R. Soc. A., 375, 20170219 (https://doi.org/10.1098/rsta.2017.0219)

\bibitem[\protect\citeauthoryear{Henyey et al.}{1965}]{Henyey1965} Henyey, L., Vardya, M.~S., \& Bodenheimer, P.\ 1965, \apj, 142, 841

\bibitem[\protect\citeauthoryear{Herwig}{2000}]{Herwig2000} Herwig, F.\ 2000, \aap, 360, 952


\bibitem[\protect\citeauthoryear{Kippenhahn et al.}{1980}]{Kippenhahn1980} Kippenhahn, R., Ruschenplatt, G., \& Thomas, H.-C.\ 1980, \aap, 91, 175

\bibitem[\protect\citeauthoryear{K\"onyves-T\'oth et al.}{2020}]{Konyves2020} K\"onyves-T\'oth, R. and Vink\'o, J., Ordasi, A., S\'arneczky, K., B\'odi, A. et al.\  2020, ApJ, 892, 121K 

\bibitem[\protect\citeauthoryear{Kurucz \& Bell}{1995}]{Kuruz1995} Kurucz R. L., Bell B. \ 1995, Atomic line list


\bibitem[\protect\citeauthoryear{Langer et al.}{1985}]{Langer1985} Langer, N., El Eid, M.~F., \& Fricke, K.~J.\ 1985, \aap, 145, 179

\bibitem[\protect\citeauthoryear{Maoz}{2014}]{Maoz2014} Maoz, D., Mannucci, F., \& Nelemans, G. \ 2014, ARA\&A, 52, 107

\bibitem[\protect\citeauthoryear{Muller}{2017}]{Muller2017} Mu\"ller, B. \ 2017, IAUS, 329, 17

\bibitem[\protect\citeauthoryear{Nicholl}{2020}]{Nicholl2020} Nicholl, M., Blanchard, P. K., Berger, E., Chornock, R., Margutti, R. et al. \ 2020, Nat. Astron., 4, 893

\bibitem[\protect\citeauthoryear{Ohkubo et al.}{2006}]{Ohkubo2006} Ohkubo, T., Umeda, H., Maeda, K., Nomoto, K., Suzuki, T. et al. \ 2006, Apj, 645, 1352





\bibitem[\protect\citeauthoryear{Pandey et al.}{2021}]{Pandey2021} Pandey, S. B., Kumar, A., Kumar, B., Anupama, G. C., Srivastav, S. et al. \ 2021, MNRAS, 507, 1229

\bibitem[\protect\citeauthoryear{Paxton et al.}{2011}]{Paxton2011} Paxton, B., Bildsten, L., Dotter, A., Herwig, F., Lesaffre, P. et al.\ 2011, apjs, 192, 3

\bibitem[\protect\citeauthoryear{Paxton et al.}{2013}]{Paxton2013} Paxton, B., Cantiello, M., Arras, P.,  Bildsten, L., Brown, E. F. et al.\ 2013, apjs, 208, 4

\bibitem[\protect\citeauthoryear{Paxton et al.}{2015}]{Paxton2015} Paxton, B., Marchant, P., Schwab, J., Bauer, E. B., Bildsten, L. et al.\ 2015, apjs, 220, 15

\bibitem[\protect\citeauthoryear{Paxton et al.}{2018}]{Paxton2018} Paxton, B., Schwab, J., Bauer, E.~B., Bildsten, L., Blinnikov, S. et al.\ 2018, apjs, 234, 34

\bibitem[\protect\citeauthoryear{Paxton et al.}{2019}]{Paxton2019} Paxton, B., Smolec, R., Schwab, J., Gautschy, A., Bildsten, L. et al.\ 2019, apjs, 243, 10

\bibitem[\protect\citeauthoryear{Piran et al.}{2019}]{Piran2019} Piran, Tsvi, Nakar, Ehud, Mazzali, Paolo, \& Pian, Elena \ 2019, ApJL, 871, L25 

\bibitem[\protect\citeauthoryear{Smartt}{2009}]{Smartt2009} Smartt S. J.\ 2009, ARA\&A, 47, 63

\bibitem[\protect\citeauthoryear{Stabrowski} {1997}]{Stabrowski1997} Stabrowski M. M. 1997, Simul. Modelling Pract. Theory, 5, 333   

\bibitem[\protect\citeauthoryear{Sukhbold et al.}{2016}]{Sukhbold2016} Sukhbold, T., Ertl, T., Woosley, S. E., Brown, Justin M., \& Janka, H. -T \ 2016, ApJ, 821, 38

\bibitem[\protect\citeauthoryear{Swartz, Sutherland \& Harkness}{1995}]{Swartz1995} Swartz D. A., Sutherland P. G., Harkness R. P. 1995, ApJ, 446, 766

\bibitem[\protect\citeauthoryear{VanRossum}{2016}]{VanRossum2016} Van Rossum, D. R., Kashyap, R., Fisher, R., et al. 2016, ApJ, 827, 128

\bibitem[\protect\citeauthoryear{von Neumann \& Richtmyer}{1950}]{Von1950} von Neumann J., Richtmyer R. D. 1950, J. Appl. Phys., 21, 232 

\bibitem[\protect\citeauthoryear{Verner, Verner \& Ferland }{1996}]{Verner1996} Verner D. A., Verner E. M., Ferland G. J. 1996, At. Data Nucl. Data Tables, 64, 1

\bibitem[\protect\citeauthoryear{Whelan}{1973}]{Whelan1973} Whelan, J., \& Iben, I., Jr. 1973, ApJ, 186, 1007


\bibitem[\protect\citeauthoryear{Woosley \& Janka}{2005}]{Woosley2005} Woosley, S. E., \& Janka, T.\ 2005, Nature Physics, 1, 147

\bibitem[\protect\citeauthoryear{Woosley}{2017}]{Woosley2017} Woosley, S. E.\ 2017, ApJ, 836, 244



\end{thebibliography}



\end{document}